\documentclass[12pt,preprint]{aastex}
\usepackage{amsmath}                
\usepackage{amsfonts}               
\usepackage{amssymb}                
\def \AU{~\rm{AU}}

\begin{document}

\title{FURTHER INDICATIONS AGAINST JET ROTATION IN YOUNG STELLAR OBJECTS}

\author{Noam Soker\altaffilmark{1}}

\altaffiltext{1}{Department of Physics, Technion$-$Israel
Institute of Technology, Haifa 32000 Israel;
soker@physics.technion.ac.il.}

\begin{abstract}
I discuss recent observations of asymmetries in Doppler shifts across T Tauri jets,
and argue that the observed asymmetric velocity shifts and gradients
do not indicate jet rotation.
These observations, therefor, cannot be used as a support of a magnetized disk wind.
The interaction of the jets with a twisted-tilted (wrapped) accretion
disk (or the variable velocity precessing model) accounts better for the observations.
\end{abstract}

\section{INTRODUCTION}
\label{sec:intro}
Asymmetry in the line of sight velocity across some jets ejected by
young stellar objects (YSOs) has been interpreted as caused by
a large scale rotation of the material in each jet around the jet's
symmetry axis (e.g., Bacciotti et al.\ 2002; Coffey et al.\ 2004;
Coffey et al. 2007; hereafter B2002, C2004 and C2007).
Rotation around the jet axis is predicted by the
magneto-centrifugal acceleration (MCA) model for jet launching
(e.g., Anderson et al. 2003).
In these models the magnetic fields that are anchored into the
accretion disk-star system play a dominate role in accelerating
the jet's material from the accretion disk.

In an earlier paper (Soker 2005) I analyzed the results of B2002 and C2004 and argued
that the observations do not support the earlier interpretation of jets
rotating around their symmetry axes. Instead, I proposed that
interaction of the jets with a twisted-tilted (wrapped) accretion
disk can form the observed asymmetry in the jets' line of sight
velocity profiles.
The jets interact with the ambient gas residing on the two sides of the disk,
e.g., a weak disk outflow or corona.
The proposed scenario is based on two plausible assumptions.
\begin{enumerate}
\item There is an inclination between the jet and the
outer parts of the disk, such that the jet is perpendicular to the inner
part of the disk; i.e., there is a twisted-tilted (wrapped)
disk.
\item The disk-jet interaction slows the jet down as the jet
entrains mass from the disk, with larger deceleration of jet
segments closer to the tilted disk.
\end{enumerate}
I showed there that the proposed scenario can
account for the basic properties of the observed velocity
profiles, while offering the advantage of not needing to
refer to any magnetic jet launching model, and there is no need to
invoke jet rotation with a huge amount of angular momentum.

The claim that the observations of B2002 and C2004 do not support jet rotation
was strengthened by the numerical simulations of Cerqueira et al. (2006).
They assumed a precessing jet whose ejection velocity changes periodically with a period
equals to the precession period.
Practically, the dependance of the jet's expansion velocity on direction around the
symmetry axis leads to the same effect as the model of Soker (2005).
Whereas in Soker (2005) the physical process behind this velocity profile is an interaction
with the material in the jet's surroundings, Cerqueira et al. (2006) give no justification
for the periodic variation of the jet's ejection speed.
Therefore, as far as comparison with observation is considered, it is hard to distinguish between
the model of jet interaction with its surrounding (Soker 2005), and the periodic jet's speed
of Cerqueira et al. (2006).
I will list them together in the next section.

\section{COMPARING OBSERVATIONS WITH MODELS}
\label{compare}

In this section I list some observations, and compare them with the
two proposed explanations for the observed (B2002, C2004 and C2007)
asymmetric velocity profiles in jets from YSO:
The interaction of the jet with the ambient gas coming from the disk (Soker 2005; as mentioned
above the model of  Cerqueira et al. 2006 is basically the same),
and with the MCA as was used by B2002, C2004 and C2007 (e.g., Anderson et al. 2003).
These are summarized in Table \ref{tab1}.

The first four points were discussed in detail in the appendix of Soker
(2005-on line material in A\&A), and therefore I summarize them here briefly.
The last four points are new.

\begin{enumerate}
\item
{\bf Stochastic velocity profiles.}
Different lines show very different velocity profiles. These
differences are larger than the claimed rotational velocity of the
corresponding jets. The interaction of the jet with its
surrounding gas is expected to be somewhat stochastic, leading to
different regions having different properties, like velocity and
density, and hence different lines are prominent in different
regions (Soker 2005).
\item
{\bf Angular momentum.} If the radial velocity gradient is
attributed to the jet's rotation, then the specific angular momentum of
the gas at the edge of the jets is extremely high.
This implies (see Soker 2005) that if C2004 interpretation of the jets in
RW Aur is correct, then according to the jet-disk models they use, the disk
must lose {\it all} of its angular momentum already at $r_d=0.4 \AU$.
This is a much larger radius than that of the accreting central star,
which implies that the disk is truncated at a very large radius,
where escape speed is much below the jet's speed.
\item
{\bf Energy.} In the analysis of Anderson et al.\ (2003;
see also 2005), the poloidal speeds along jets' stream lines of
the YSO DG Tauri are $<3$ times what would be the escape velocity
at the footpoints of the stream lines.  The velocities
found by C2004 are much higher than the theoretical expectation.
In the YSO LkH$\alpha$ 321, for example, the jet speed is $\sim 4-4.5$
times the escape velocity.
In one streamline in DG Tau this ratio as given by C2007 is $\sim 5$.
The required efficiency of
energy transfer from accreted to ejected mass seems to be higher
than that expected in MCA models.
Higher velocities are obtained in the numerical simulations of
Garcia et al.\ (2001) which include the thermal state of the jet,
but the fraction of ejected mass (out of the accreted mass) is $1-2 \%$,
much below the value of $\sim 10 \%$ in RW Tau (Woitas et al.\ 2005).
\item
{\bf Model versus observations.}
C2004 use the jet launching model as presented by Anderson et al.\
(2003), who use it for the YSO DG Tau. However, in that model the
toroidal velocity decreases with distance from the jet's axis,
while in the observations, both of DG Tau (Anderson et al.\ 2003)
and the three YSOs studied by C2004, the toroidal velocity
increases with distance from the jet's axis.
Pesenti et al. (2004) present a more detailed MHD model based on
jet rotation to account for the observation of DG Tau (B2002).
Pesenti et al. (2004) find a good fit between their model and the
velocity map of DG Tau. However, the model of Pesenti et al. (2004) for
DG Tau has two significant differences from that of TW Aur (Woitas et al.\ 2005).
First, their derive typical toroidal velocity for stream lines with footpoints of
$\sim 1 \AU$ is much lower than that in TW Aur.
The second difference involved the theoretical model used by
Pesenti et al. (2004). Pesenti et al.\ (2004) find  that only the
warm jet model fit the observations. In this model the jet is
thermally-driven (Casse \& Ferreira 2002). Both these differences,
that the jets have low specific angular momentum, and are
thermally driven, are along the main theme of the present paper,
although significant differences exist between the warm MHD model
used by Pesenti et al.\ (2004) and the model proposed by Soker \& Regev (2003).
Because a more sophisticated and detailed MHD models for jet rotation
might overcome these problems, I put `$=$' for the MCA model in Table \ref{tab1}.
\item
{\bf Faint low velocity component.}
In the position-velocity maps of CW Tau in [OI]~$\lambda 6300$
and of DG TAu and TH 28 in the Mg~II line, there is a faint component
of velocity close to zero, or even with a velocity opposite to that of the jet
(C2007).
Such a component is expected from the backflowing material formed from the jet-ambient
medium interaction, e.g., Stone \& Norman (1993, figs. 13 and 16) and (1994, fig 8).
This component has no particular explanation in the MCA model.
\item
{\bf Counter disk rotation. }
Cabrit et al. (2006) found the rotation of the disk in RW Aur to be
in opposite direction to the jet rotation argued for by Woitas et al. (2005).
This is in contradiction with the MCA model.
\item
{\bf No velocity gradient in HH~30.}
The jet of HH~30 is in the plane of the sky (inclination of $1^\circ$).
At that inclination the jet rotation should have its larger amplitude.
But there is no indication for jet rotation in HH~30 (C2007),
or in the CO outflow associated with it (Pety et al. 2006).
In the jet interaction model at that inclination the velocity gradient should
be zero.
TH~28 is also close to the plane of the sky (inclination of $10^\circ$),
and show very low velocity asymmetries (C2007).
\item
{\bf Asymmetry between two opposite jets.}
In TH~28 the two jets were observed by C2007. the velocity profiles
of the two jets are dissimilar. In the jet-ambient medium interaction this is
accounted for by a different surrounding medium on the two sides.
If the velocity profiles were due to jets' rotation, the two jets
should have shown similar velocity profiles.
This observation pose a problem to the periodic jet's speed model
of Cerqueira et al. (2006) as well.
\end{enumerate}

\begin{table}
\begin{minipage}[t]{\columnwidth}
\caption{Observations and theory}
\label{tab1}
\centering
\renewcommand{\footnoterule}{ }  
\begin{tabular}{lll}
\hline \hline
 Observed property  & Jet-disk interaction  & Magneto-centrifugal   \\
                    & (Soker 2005)        & acceleration  \\
\hline \hline
(1) Different lines show   & (+)\footnote{(+) Expected by the model; ($-$) Cannot (very difficult to)
be explained by the model; ($=$) Possible to explain by the model}
                           Interaction with disk  & ($=$) Possible to explain     \\
very different velocity                  & forms different regions     &  under a particular jet      \\
profiles.       & with different properties.      &  structure only.       \\
\hline
(2) If the velocity profile   & (+) The velocity gradients  & ($-$) The disk loses most  \\
is rotation, then jet's        & do not indicate rotation.   & of its angular momentum    \\
angular momentum is large.    &                              & at large radius.        \\
\hline
(3) Observed jets' outflow & (+) Jet are launched from the  & ($-$) Too high for the MCA      \\
speed.                     & inner radius of the disk, where      &   model.     \\
                           & escape velocity is high enough.   & \\
\hline
(4) Rotation interpretations   & (+) In the thermal-launching  & ($=$) The model used to explain    \\
    do not fit well the         & model only a minor rotation      &  observations is not in accord      \\
    model used.                   & is expected.                   &  with the observations.       \\
  \hline
(5) A separate low velocity  & (+) Emission from the cocoon  &  ($-$) No explanation.     \\
component.                 &   or from the interacting    &    \\
                                   & surroundings.           &    \\
\hline
(6) Disk and jet rotations  & (+) The observations are not  & ($-$) In contradiction.    \\
    in RW Aur are opposite.       &  of jet's rotation.  &    \\
\hline
(7) No rotation signature  & (+) No velocity gradient  & ($-$) Rotation signature    \\
in HH~30, although the jet     &  is expected.            & should be at maximum.        \\
is in the plane of the sky. &      &        \\
\hline
(8) The two opposite jets  & (+) The ambient medium            & ($-$) Not expected if velocity    \\
of TH 28 show different   &  is different in the           &   gradients are due to rotation.     \\
velocity profiles.         &   two sides.                  &        \\
\hline

\hline
\end{tabular}
\end{minipage}
\end{table}
%

\section{SUMMARY}
\label{summary}
This short paper has one point (Soker 2005; Cerqueira et al. 2006):
The observed asymmetric velocity shift on the two sides of some YSO jets
cannot serve as confirmation for the magneto-centrifugal acceleration (MCA)
model for jet formation.
In this paper I went further, and argue that the observed velocity gradients
and asymmetries are actually in contradiction with the MCA model.

This research was supported by a grant from the
Asher Space Research Institute at the Technion.


\begin{references}

\reference{} Anderson, J. M., Li, Z.-Y., Krasnopolsky, R., \&
    Blandford, R. D. 2003, ApJ, 590, L107

\reference{} Anderson, J. M., Li, Z.-Y., Krasnopolsky, R., \&
    Blandford, R. D. 2005, ApJ, 630, 945

\reference{} Bacciotti, F., Ray, T. P.,  Mundt, R., Eisl\"offel, J., \& Solf, Jo.
    ApJ, 576, 222 (B2002)

\reference{} Cabrit, S., Pety, J., Pesenti, N., \& Dougados, C. 2006, A\&A, 452, 897

\reference{} Casse, F., \& Ferreira, J. 2000, A\&A, 361, 1178

\reference{} Cerqueira, A. H., Velazquez, P. F., Raga, A. C., Vasconcelos, M. J.,
     \& de Colle, F. 2006, A\&A, 448, 231

\reference{} Coffey, D., Bacciotti, F., Woitas, J., Ray, T. P.,
           \& Eisloffel, J. 2004, ApJ, 604, 758 (C2004)

\reference{} Coffey, D., Bacciotti, F., Ray, T. P.,  Ray, T. P., Eisloffel, J., \&
            Woitas, J., (astro-ph/0703271) (C2007)

\reference{} Garcia, P. J. V., Ferreira, J., Cabrit, S., \& Binette, L. 2001, A\&A, 377, 589

\reference{} Pesenti, N., Dougados, C., Cabrit, S., Ferreira, J.,
  Casse, F., Garcia, P., \&  O'Brien, D. 2004, A\&A, 416, L9

\reference{} Pety, J., Gueth, F., Guilloteau, S., \&  Dutrey, A. 2006, A\&A, 458, 841

\reference{} Soker, N. 2005, A\&A, 435, 125

\reference{} Soker, N., \& Regev, O. 2003, A\&A, 406, 603

\reference{} Stone, J. M., \& Norman, M. L. 1993, ApJ, 413, 210

\reference{} Stone, J. M., \& Norman, M. L. 1994, ApJ, 420, 237

\reference{} Woitas, J., Bacciotti, F., Ray, T. P., Marconi, A.,
  Coffey, D., \& Eisloffel, J. 2005, A\&A, 432, 149



\end{references}
\end{document}